\documentclass{amia}
\usepackage{lipsum} 
\usepackage{xcolor} 
\usepackage{hyperref}
\usepackage{amsmath}
\usepackage{amsfonts}
\usepackage{bbm}
\usepackage{algorithm}
\usepackage{algpseudocode}

\newcommand{\logit}{\operatorname{logit}} 
\DeclareMathOperator{\IDF}{\mathbbm{1}}
\DeclareMathOperator{\Prob}{\mathbb{P}}

\begin{document}

\title{powerROC: An Interactive Web Tool for Sample Size Calculation in Assessing Models' Discriminative Abilities}

\author{François Grolleau, MD, PhD$^1$, Robert Tibshirani, PhD$^{2,3}$, Jonathan H. Chen, MD, PhD$^{1,4,5}$ }

\institutes{
    $^1$Stanford Center for Biomedical Informatics Research, Stanford University \\
    $^2$Department of Statistics, Stanford University \\
    $^3$Department of Biomedical Data Science, Stanford University \\
    $^4$Stanford Clinical Excellence Research Center, Stanford University \\
    $^5$Division of Hospital Medicine, Stanford University \\
}

\maketitle

\section*{Abstract}

\textit{Rigorous external validation is crucial for assessing the generalizability of prediction models, particularly by evaluating their discrimination (AUROC) on new data. This often involves comparing a new model's AUROC to that of an established reference model. However, many studies rely on arbitrary rules of thumb for sample size calculations, often resulting in underpowered analyses and unreliable conclusions. This paper reviews crucial concepts for accurate sample size determination in AUROC-based external validation studies, making the theory and practice more accessible to researchers and clinicians. We introduce powerROC, an open-source web tool designed to simplify these calculations, enabling both the evaluation of a single model and the comparison of two models. The tool offers guidance on selecting target precision levels and employs flexible approaches, leveraging either pilot data or user-defined probability distributions. We illustrate powerROC’s utility through a case study on hospital mortality prediction using the MIMIC database.}

\section*{Introduction}

In medicine, the abundance of prediction models—with hundreds existing for conditions like chronic obstructive pulmonary disease \cite{bellou2019prognostic}, cardiovascular disease \cite{damen2016prediction}, and COVID-19 \cite{wynants2020prediction}—stands in stark contrast to the scarcity of their validation \cite{van2023there}. This disparity poses a significant challenge, as a model's true value lies in its ability to generalize beyond its developmental cohort. External validation is crucial for demonstrating this generalizability; however, many validation studies are underpowered, yielding unreliable results and potentially misleading conclusions \cite{collins2014external, groot2021availability}.

Proper sample size calculation is a critical, yet often overlooked, aspect of effective validation \cite{riley2024evaluation}. Researchers often rely on rules of thumb derived from simulation and resampling studies. For instance, a common suggestion is to include at least 100 events and 100 non-events to precisely estimate the area under the receiver operating characteristic curve (AUROC) \cite{vergouwe2005substantial, collins2016sample}. While these heuristics provide a starting point, they can lead to sample sizes that are either too small, compromising statistical power, or unnecessarily large, resulting in excessive data collection that is time-consuming and expensive. A more rigorous approach not only ensures efficient resource allocation but also minimizes the risk of type II errors—incorrectly concluding no significant difference in models' discriminative abilities when one truly exists.

Recognizing the importance of proper sample size determination, international reporting guidelines such as the TRIPOD+AI statement now mandate its justification in validation studies \cite{collins2024tripod+}. While software for evaluating model performance exists \cite{Ensor2023, goksuluk2016easyroc, robin2011proc}, most available tools neglect sample size calculation for comparing two prediction models or lack an accessible Web interface, limiting their practical application. Table \ref{tab:softwares} summarizes the key differences between these existing packages.

To address this gap, we present powerROC: an interactive Web tool designed to assist researchers in determining the sample size required for an external validation study. Our objective is to make the theory and practice of sample size calculation for external validation studies more accessible by providing a clear review of key AUROC concepts and relevant references as well as a user-friendly web interface. We illustrate the utility of powerROC through a case study focused on hospital mortality prediction using the MIMIC database. The web tool is freely available at \url{https://fcgrolleau.github.io/powerROC/}, with an introduction tutorial and the source code for implementation and case study on the MIMIC database accessible at \url{https://github.com/fcgrolleau/powerROC}.

\begin{table}[h!]
\caption{Software available for calculating the sample size required for an external validation study}\label{tab:softwares}
\begin{tabular}{lccccc}
\hline\hline
                                                                              & \multicolumn{1}{l}{pmvalsampsize} & \multicolumn{1}{l}{easyROC} & \multicolumn{1}{l}{pROC} & \multicolumn{1}{l}{powerROC} \\ \hline
Sample size for evaluating a single prediction model                          & Yes                               & Yes                         & Yes & Yes                          \\
Sample size for comparing two prediction models                               &                                   &                             &     &                            \\
$\,$ - With a pilot test set                                                  & No                                & No                          & Yes & Yes                          \\
$\quad$ Allows to vary the prevalence                                         & —                                 & —                           & No  & Yes                          \\
$\,$ - Without a pilot test set                                               & No                                & Yes                         & Yes & Yes                          \\
$\quad$ Accounts for the paired study design                                  & —                                 & No                          & Yes & Yes                          \\
$\quad$ Interpretability of the parameters                                    & —                                 & No                          & No  & Yes                          \\
$\quad$ Visualization of the probability distribution specified               & —                                 & No                          & No  & Yes                          \\
Web interface                                                                 & No                                & Yes                         & No  & Yes                          \\
Programming language                                                          & R \& Stata                        & R                           & R   & Python                       \\ \hline\hline
\end{tabular}
\end{table}

\section*{Methods}
\subsection{Methods for AUROC estimation and comparison}\label{subsec:methods}
\paragraph{Definition of AUROC}
The Area Under the Receiver Operating Characteristic curve (AUROC) is a widely used performance metric for binary outcome prediction models. It quantifies a model's discriminative ability—its capacity to distinguish between individuals with the outcome (cases) and those without (controls). The AUROC ranges from 0.5 to 1, where 1 indicates perfect discrimination and 0.5 suggests the model performs no better than chance. More precisely, the AUROC represents the probability that, for any randomly selected pair of participants (one case and one control), the model assigns a higher risk to the case. Denoting the true unknown value of AUROC as $\theta$, we can express this as:

\begin{equation*}
\theta = \Prob(\Tilde{Y}_1^x > \Tilde{Y}_0^x)
\end{equation*}

where $\Tilde{Y}_1^x$ and $\Tilde{Y}_0^x$ are the predictions from classifier $x$ for cases and controls, respectively \cite{krzanowski2009roc}.

\begin{figure}[h!]
    \centering
    \includegraphics[width=1\linewidth]{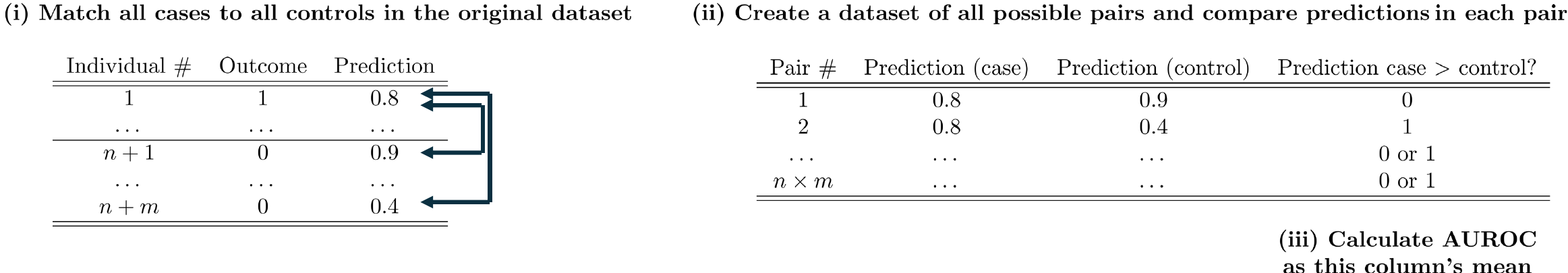}
    \caption{Non-parametric procedure to estimate AUROC. With $n$ cases and $m$ controls, there are $n \times m$ pairs to consider.}
    \label{fig:est_procedure}
\end{figure}

\paragraph{Estimation procedure for AUROC}
The standard approach to estimate the AUROC from data involves considering all possible pairs of cases and controls and calculating the proportion of pairs where the model assigns a higher risk to the case. This procedure is illustrated in Figure \ref{fig:est_procedure}.

This estimation procedure can be succinctly described as:

\begin{equation}\label{eq:auroc_estimator}
\widehat{\theta}=\frac{1}{nm}\sum_{i=1}^n \sum_{j=1}^m \mathbbm{1} \big\{\Tilde{Y}_{1,i}^x > \Tilde{Y}_{0,j}^x \big\}
\end{equation}

where $\Tilde{Y}_{1,i}^x, \Tilde{Y}_{0,j}^x$ are the predictions from classifier $x$ for cases $i=1,\dots,n$ and controls $j=1,\dots,m$ respectively, and $\IDF\{E\}$ is the indicator function such that $\IDF\{E\} = 1$ if the event $E$ is true and $= 0$ otherwise \cite{hanley1982meaning}.

\paragraph{Asymptotic properties of the AUROC estimator}
The AUROC estimator in equation \ref{eq:auroc_estimator} is recognized as a particular type of U-statistic\footnote{Informally, a U-statistic is an estimator that takes the form of an average of a (symmetric) function applied to subsets (e.g., pairs, triples, or larger groups) of observations from a sample. See van der Vaart, Chapter 12 \cite{van2000asymptotic} for a full introduction to U-statistics.} (specifically, the Mann-Whitney U-statistic) \cite{bamber1975area}. Hoeffding \cite{hoeffding1948class} demonstrated that U-statistics are consistent and asymptotically normal. Therefore, as the sample size $N=n+m$ increases, the U-statistic $\widehat{\theta}$ approximately follows a normal distribution with mean $\theta$ (the true value of the AUROC) and its variance approaches zero. Newcombe \cite{newcombe2006confidence} showed that this variance can be expressed as:

\begin{equation}\label{eq:var_auroc}
\text{VAR}(\widehat{\theta})=\frac{\theta(1-\theta)\Big(1+\big(\frac{N}{2}-1\big)\big(\frac{1-\theta}{2-\theta}\big)+\frac{(\frac{N}{2}-1)\theta}{1+\theta}\Big)}{\phi (1-\phi) N^2}   
\end{equation}

where $\phi=\frac{n}{N}$ denotes the prevalence. This result enables straightforward calculation of asymptotic (i.e., non-bootstrap) confidence intervals (CIs) for the AUROC as $\widehat{CI}_{95\%}(\theta)=\widehat{\theta} \pm 1.96 \times \sqrt{\widehat{\text{VAR}}(\widehat{\theta})}$. More importantly, this procedure can be reversed to calculate the sample size required to estimate the AUROC with a desired level of precision (see Methods, section \ref{subsec:single_model}).

\paragraph{DeLong p-values for comparing AUROCs from two prediction models}
The properties of U-statistics can be exploited to test the null hypothesis that two models have the same AUROC, i.e., $H_0: \, \theta^A = \theta^B$ where $\theta^A$ and $\theta^B$ are the true AUROCs from two classifiers denoted $A$ and $B$ respectively. Under the null hypothesis $H_0$, the variable
\begin{equation}\label{eq:z_test}
    Z=\frac{\widehat{\theta}^A - \widehat{\theta}^B}{\sqrt{\text{VAR}(\widehat{\theta}^A - \widehat{\theta}^B)}}=\frac{\widehat{\theta}^A - \widehat{\theta}^B}{\sqrt{\text{VAR}(\widehat{\theta}^A) + \text{VAR}(\widehat{\theta}^B)- 2\text{COV}(\widehat{\theta}^A,\widehat{\theta}^B)}}
\end{equation}

approximately follows a standard normal distribution. When predictions from models $A$ and $B$ are obtained from the same individuals (paired study design), the covariance term $\text{COV}(\widehat{\theta}^A,\widehat{\theta}^B)$ is non-zero and typically positive. DeLong et al. \cite{delong1988comparing} proposed a procedure to simultaneously estimate $\text{VAR}(\widehat{\theta}^A)$, $\text{VAR}(\widehat{\theta}^B)$, and $\text{COV}(\widehat{\theta}^A,\widehat{\theta}^B)$. Substituting these estimated values into equation \ref{eq:z_test} allows calculation of DeLong p-values to test the null hypothesis $H_0$. Furthermore, by specifying a relevant alternative hypothesis, these p-values can be utilized to calculate the sample size required for comparing two prediction models with a desired level of statistical power (see Methods, section \ref{subsec:two_models}).

\subsection{Calculating the necessary sample size to evaluate a single prediction model}\label{subsec:single_model}
Accurate evaluation of a prediction model's discriminative performance for a binary outcome requires precise assessment of the AUROC. To determine the sample size needed for such an evaluation, we propose a method that inverts the formula used for calculating AUROC confidence intervals (Equation \ref{eq:var_auroc}). Our approach takes three key inputs: the anticipated AUROC, the prevalence of the outcome in the evaluation population, and a target width for the estimated AUROC 95\% confidence interval. We use an iterative process (outlined in Algorithm \ref{alg:1}) to determine the minimum sample size required to achieve the desired precision, systematically increasing the sample size until the calculated standard error (used to construct the confidence interval) meets or exceeds the target level of precision. To balance statistical precision with practical feasibility we recommend using a target confidence interval width of $\leq 0.1$. 

\begin{algorithm}[!h]
    \caption{Calculate the sample size to precisely estimate the AUROC of a single prediction model}\label{alg:1}
\begin{algorithmic}[1]
    \Require $\theta:$ anticipated AUROC, $\phi:$ prevalence, $l:$ target width for the estimated AUROC 95\% confidence interval
    \State calculate the target standard error $se_t \gets \frac{l}{2 \times 1.96}$
    \For{$N = 1, 2, \dots $}
        \State Calculate a standard error for the current sample size: $$se_N \gets \sqrt{\frac{\theta(1-\theta)\Big(1+\big(\frac{N}{2}-1\big)\big(\frac{1-\theta}{2-\theta}\big)+\frac{(\frac{N}{2}-1)\theta}{1+\theta}\Big)}{\phi (1-\phi) N^2}}$$
        \If{$se_N < se_t$}
            \State \Return Sample size $N$
        \EndIf
    \EndFor
\end{algorithmic}
\end{algorithm}

\subsection{Calculating the necessary sample size to compare two prediction models}\label{subsec:two_models}
In the rapidly evolving field of machine learning applied to healthcare, new classifiers for prognosis or diagnosis are frequently developed, often claiming superior performance over existing models. A critical aspect of evaluating these claims involves determining whether the new model achieves greater discriminative ability than an established reference model. This evaluation typically requires conducting an external validation study and comparing the AUROCs of the new and reference models, utilizing DeLong's test (See Methods, section \ref{subsec:methods}). A crucial consideration in designing such a comparative study is determining the sample size necessary to achieve a prespecified level of statistical power at a given significance threshold ($\alpha$). To achieve this, we need a precise specification of the alternative hypothesis, which outlines how the true AUROCs of the new and reference models are expected to differ. It is important to note that simply specifying the difference between the two true AUROCs is insufficient for this purpose. We propose two approaches to provide a precise specification of the alternative hypothesis: (i) utilizing a pilot test set, or (ii) having the user specify a probability distribution. Both methods leverage DeLong p-values and Monte Carlo simulations to determine the minimum sample size required to detect statistically significant differences in AUROCs with a desired level of power.

\paragraph{With a pilot test set}
When researchers have access to a pilot dataset, potentially small, representative of the future external validation study, this dataset can be used to estimate the power that a given sample size would yield for comparing the AUROCs of two prediction models. This pilot dataset must include labels and predictions from both models ($A$ and $B$).
Our approach leverages nonparametric Monte Carlo simulations to estimate power and is described in detail in Algorithm \ref{alg:2}. Briefly, the procedure involves resampling with replacement from the pilot dataset to generate numerous datasets of a larger (or smaller) size than the original. For each resampled dataset, we calculate a p-value from DeLong's test for paired data. Power is then estimated as the proportion of p-values below a specified threshold. For practical feasibility and statistical precision, we recommend selecting the sample size that achieves 80\% power at the $\alpha$ threshold of 0.05. If the prevalence in the external validation study is expected to differ from that of the pilot dataset, Algorithm \ref{alg:2} can be modified to account for this; the user can input the anticipated prevalence, and the procedure will perform weighted sampling to estimate power (\emph{see} footnote below.)

\begin{algorithm}
    \caption{Using a pilot test set, compute the power that a given sample size would yield to compare the AUROCs of two prediction models}\label{alg:2}
\begin{algorithmic}[1]
    \Require $(Y_i,\tilde{Y}^A_i,\tilde{Y}^B_i)_{1\leq i\leq n}:$ pilot testset of size $n$ comprising labels, predictions from model $A$ and predictions from model $B$, $\alpha:$ significance threshold, $N:$ sample size to evaluate, $M:$ no. of simulation iterations 
    \For{$m = 1, 2, \dots, M$}
        \State Sample $N$ examples $j_1,j_2,\dots,j_N$ with replacement uniformly\footnotemark{} from $\{1,2,\dots,n\}$
        \State Calculate a p-value $\rho_m$ from DeLong test for paired data using the dataset $(Y_{j_k},\tilde{Y}^A_{j_k},\tilde{Y}^B_{j_k})_{1\leq k\leq N}$
    \EndFor
    \State \Return Power at sample size $N$ and significance threshold $\alpha$ as: $\frac{1}{M}\sum_{m=1}^M \IDF \{\rho_m < \alpha\}$
\end{algorithmic}
\end{algorithm}

\footnotetext{Please note that this approach differs from bootstrapping, which typically entails generating numerous datasets of the same size as the original. To enable users to adjust the prevalence in the test set to a specified value $\phi$, Algorithm \ref{alg:2} can be modified to use weighted sampling with replacement. The weights are defined as $ w_i = \frac{\phi Y_i}{\sum_{i=1}^n Y_i} + \frac{(1-\phi)(1-Y_i)}{\sum_{i=1}^n (1-Y_i)} $. This reweighting scheme ensures that $\sum_{i=1}^n w_i = 1$ and $\sum_{i=1}^n w_i Y_i = \phi$ thereby achieving the desired prevalence.}

\paragraph{Without a pilot test set}
When a pilot test set is unavailable, users can still estimate the required sample size by specifying the anticipated data generating process. This entails defining the joint probability distribution for the labels ($Y$) and the predicted risks from both models ($\tilde{Y}^A$, $\tilde{Y}^B$) in the evaluation population. We simplify this by decomposing the joint probability density as follows:
\begin{equation*}
p_{Y,\tilde{Y}^A,\tilde{Y}^B}(y,\tilde{y}^A,\tilde{y}^B)=p_{\tilde{Y}^A,\tilde{Y}^B|Y}(\tilde{y}^A,\tilde{y}^B|y)p_Y(y).
\end{equation*}
Our approach focuses on specifying parameters for two conditional joint densities:
\begin{itemize}
    \item $p_{\tilde{Y}^A,\tilde{Y}^B|Y}(\tilde{y}^A,\tilde{y}^B|1)$, representing the distribution of predicted risks given patients are cases  ($Y=1$);
    \item $p_{\tilde{Y}^A,\tilde{Y}^B|Y}(\tilde{y}^A,\tilde{y}^B|0)$, representing the distribution of predicted risks given patients are controls ($Y=0$).
\end{itemize}
Algorithm \ref{alg:3} details how to parameterize these densities and subsequently compute the statistical power for comparing the AUROCs of two prediction models at a given sample size. In short, we utilize bivariate normal distributions for both densities, employing a reparametrization trick to constrain user input to values between $0$ and $1$ for ease of interpretation. The prevalence, serves as the single parameter for the Bernoulli distribution, $p_Y(y)$.
With the joint density defined, our method proceeds with parametric Monte Carlo simulations, mirroring the approach used when a pilot test set is available. We generate numerous datasets of the desired sample size by sampling from the specified distribution. For each dataset, we calculate the p-value from DeLong's test for paired data to compare the AUROCs. Power is then estimated as the proportion of p-values falling below a significance threshold, typically $\alpha = 0.05$. We recommend selecting the sample size that achieves 80\% power at this significance level.

\begin{algorithm}[!h]
    \caption{Specifying a probability distribution, compute the power that a given sample size would yield to compare the AUROCs of two prediction models}\label{alg:3}
\begin{algorithmic}[1]
    \Require parameters in the $]0,1[$ interval: $(\mu^A_1,\mu^B_1):$ mean parameters for cases, $(\mu^A_0,\mu^B_0):$ mean parameters for controls, $(v^A_1,v^B_1):$ variance parameters for cases, $(v^A_0,v^B_0):$ variance parameters for controls, $(r_1,r_0):$ between model correlation parameters for cases and controls respectively. Require: $\phi:$ prevalence, $\alpha:$ significance threshold, $N:$ sample size to evaluate, $M:$ no. of simulation iterations 
    \State Reparameterize the mean parameters: $\mu^x_y \gets \logit \big(y \mu^x_y + (1-y) (1-\mu^x_y)\big),$ for $x \in \{A,B\}$ and $y \in \{0,1\}$ 
    \State Reparameterize the variance parameters: $v^x_y \gets - \ln (1-v^x_y ),$ for $x \in \{A,B\}$ and $y \in \{0,1\}$ 
    \State Reparameterize the correlation parameters:  $r_y \gets r_y \sqrt{v^A_y v^B_y },$ for $y \in \{0,1\}$
    \For{$m = 1, 2, \dots, M$}
        \State Sample $N$ examples of the variables $Y, \tilde{Y}_1,$ and $\tilde{Y}_0$ as
$$Y \sim \text{Bernoulli}(\phi), \quad 
\tilde{Y}_1 \sim \mathcal{N}\left(\begin{bmatrix}
\mu^A_1 \\
\mu^B_1
\end{bmatrix}, \begin{bmatrix}
v^A_1 & r_1 \\
r_1 & v^B_1
\end{bmatrix}\right), \quad
\tilde{Y}_0 \sim \mathcal{N}\left(\begin{bmatrix}
\mu^A_0 \\
\mu^B_0
\end{bmatrix}, \begin{bmatrix}
v^A_0 & r_0 \\
r_0 & v^B_0
\end{bmatrix}\right)$$
\State Generate the predictions as $(\tilde{Y}^A,\tilde{Y}^B) \gets Y \tilde{Y}_1 + (1-Y) \tilde{Y}_0$
\State Calculate a p-value $\rho_m$ from DeLong test for paired data using the dataset $(Y_i,\tilde{Y}^A_i,\tilde{Y}^B_i)_{1\leq i\leq N}$
    \EndFor
    \State \Return Power at sample size $N$ and significance threshold $\alpha$ as: $\frac{1}{M}\sum_{m=1}^M \IDF \{\rho_m < \alpha\}$
\end{algorithmic}
\end{algorithm}

\subsection*{Results}
\subsection*{1 \hspace{0.2cm} The powerROC web tool}
\begin{figure}[h!]
    \centering
    \includegraphics[width=1\linewidth]{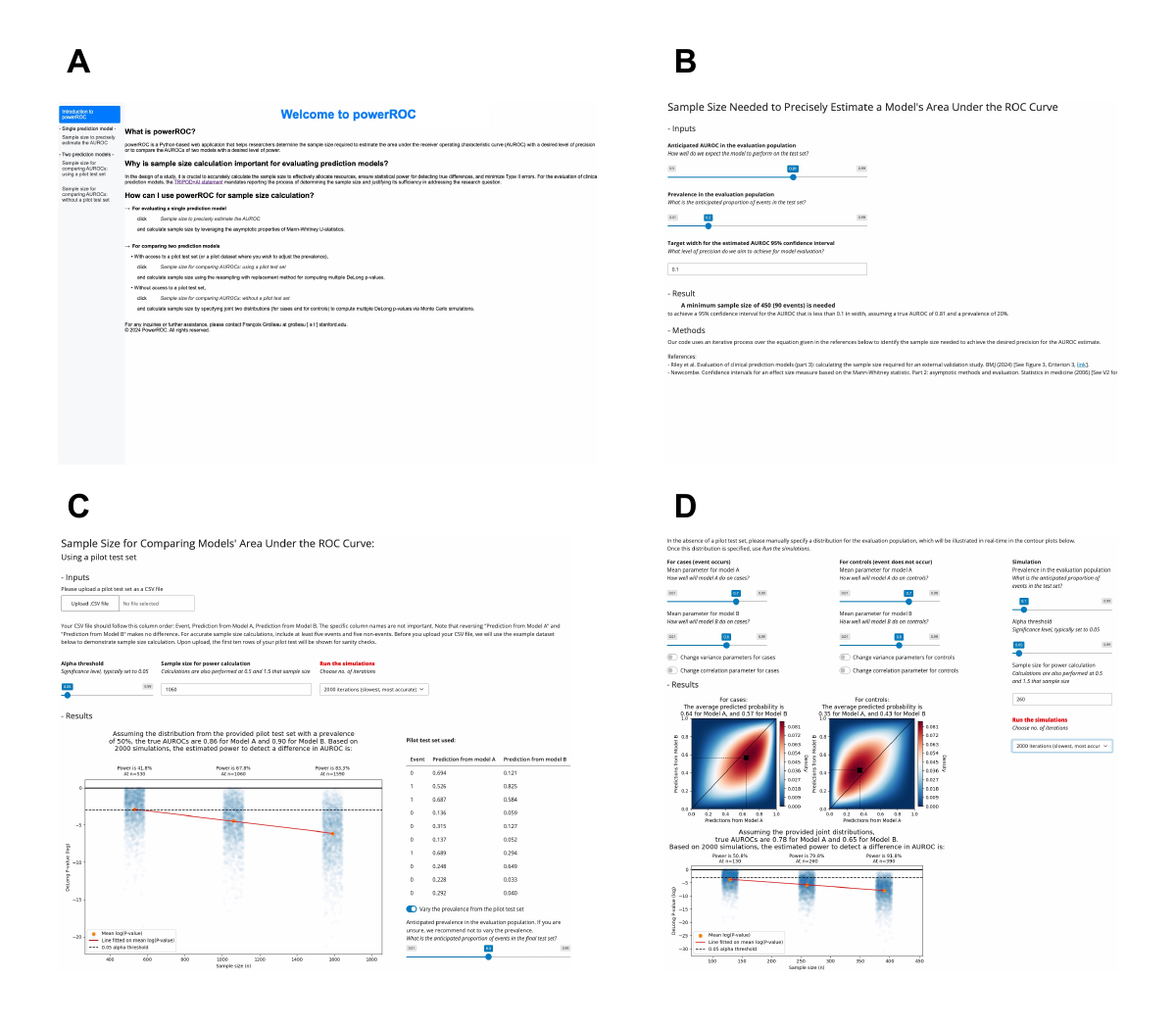}
    \caption{The powerROC web tool interface. This figure showcases the different modules within the powerROC web tool designed for sample size calculations in AUROC-based evaluations. Panel A: The Home page provides a concise overview of the tool, highlighting the importance of appropriate sample size determination for rigorous external validation of prediction models, along with navigation links to other modules. Panel B: The “Single Model AUROC” module (Algorithm \ref{alg:1}) allows users to calculate the sample size needed to precisely estimate the AUROC of a single prediction model. Panel C: The “Two Models with Pilot Data” module (Algorithm \ref{alg:2}) enables sample size estimation for comparing two models using a pilot test set (uploaded as a CSV file). Users can adjust the prevalence to reflect potential differences between the pilot and the planned external validation study. Panel D: The “Two Models - No Pilot Data” module (Algorithm \ref{alg:3}) facilitates sample size calculations when a pilot dataset is not available. Users specify parameters for a probability distribution, aided by clear descriptions and interactive contour plots. Default values for variance and correlation parameters (set to 0.9) can be modified to explore sensitivity to these parameters. Both two-model modules (Panels C and D) provide power analysis results in an accessible format, including textual interpretations and a figure summarizing the Monte Carlo simulations.}
    \label{fig:powerroc}
\end{figure}

powerROC is a freely available, open-source web tool that implements our sample size calculation methodologies described in Algorithms \ref{alg:1}, \ref{alg:2}, and \ref{alg:3}. Accessible at \url{https://fcgrolleau.github.io/powerROC/}, powerROC features an intuitive user interface (Figure \ref{fig:powerroc}), built using Shiny and Python with an HTML/JavaScript wrapper. To facilitate real-time responsiveness, Monte Carlo simulations are performed dynamically as users adjust input parameters. For faster computation of DeLong p-values, we utilize the algorithm developed by Sun et al. \cite{sun2014fast}, which achieves linearithmic time complexity with respect to sample size.

\subsection*{2 \hspace{0.2cm} Case study on mortality prediction for ICU patients}
To illustrate the practical application of powerROC, we present a case study focused on estimating sample sizes for an external validation study of a hospital mortality prediction model. While this example uses a specific model and clinical outcome, the general principles and methodologies presented here are applicable to a wide range of prediction tasks and clinical settings, including both diagnosis and prognosis prediction models.

The Super ICU Learner Algorithm (SICULA) is an ensemble machine learning model designed to predict hospital mortality in ICU patients \cite{pirracchio2015mortality}. Trained on the MIMIC-II dataset (24,508 admissions from Beth Israel Deaconess Medical Center, Boston, MA), SICULA attained an AUROC of 0.85 (95\% CI: 0.84-0.85) in cross-validation. The study also reported a lower AUROC of 0.78 (95\% CI: 0.77-0.78) for the established SAPS-II model, commonly used in clinical settings. However, rigorous external validation is needed to assess SICULA's generalizability across different hospitals, patient populations, and evolving clinical practices. This case study demonstrates how powerROC can be used to estimate sample sizes for such an external validation, considering two key scenarios: 1) precisely estimating SICULA's AUROC in a new population and 2) comparing SICULA's performance against the SAPS-II reference model.

\paragraph{Sample size needed to precisely estimate SICULA's AUROC}
To determine the sample size required to precisely estimate SICULA's AUROC in a new population, we make conservative assumptions based on potential shifts in patient characteristics. We assume a mortality prevalence of 20\% (higher than the 11\% observed in MIMIC-II) and an AUROC of 0.81 (5\% lower than reported in \cite{pirracchio2015mortality}). Targeting a 95\% confidence interval width of 0.1 for the AUROC, powerROC's implementation of Algorithm \ref{alg:1} suggests a minimum sample size of 450 patients (90 mortality events).

\paragraph{Sample size needed to compare the AUROCs of SICULA and SAPS-II}
To estimate the sample size required to compare SICULA's performance against the SAPS-II reference model, we consider scenarios with and without access to a pilot test set.
\begin{itemize}
    \item With a pilot test set: To illustrate the scenario where researchers have access to a pilot dataset, we reproduce the SICULA model by training a SuperLearner model \cite{van2007super} on the MIMIC-III database, following the methodology described in the original study. We then use patients from the MIMIC-IV database \cite{johnson2023mimic} admitted after 2013 (and therefore not included in the MIMIC-III data used for training) as a pilot test set. In this pilot set, SICULA and SAPS-II achieved AUROCs of 0.82 (95\% CI: 0.82-0.83) and 0.80 (95\% CI: 0.79-0.81), respectively. Assuming a mortality prevalence of 20\% in our hypothetical external validation study (higher than the 12\% observed in MIMIC-IV), and using 2000 simulation iterations, powerROC's  implementation of Algorithm \ref{alg:2} indicates that a minimum sample size of 590 patients (110 mortality events) is needed to achieve 80\% power for detecting a statistically significant difference in AUROCs at $\alpha = 0.05$.
    \item Without a pilot test set: When a pilot dataset is unavailable, we can leverage expert opinion to define the anticipated data distribution. Assuming a mortality prevalence of 20\%, we set the average predicted mortality probability to 0.17 for both models on controls. For cases, we specify an average predicted probability of 0.44 for SICULA and 0.41 for SAPS-II, reflecting an anticipated difference in performance. Using these specifications with default values for variance (0.9) and correlation (0.9) parameters, the anticipated AUROCs are 0.81 for SICULA and 0.78 for SAPS-II. With these inputs and 2000 simulation iterations, powerROC's implementation of Algorithm \ref{alg:3} determines that a sample size of 770 patients (154 mortality events) would be needed to achieve 80\% power at $\alpha = 0.05$.
\end{itemize}

\subsection*{Discussion}
This paper reviewed crucial concepts for accurate sample size determination in AUROC-based external validation studies and introduced powerROC, an open-source web tool designed to simplify these calculations. powerROC addresses a critical need in healthcare, where rigorous external validation is paramount to determine if a new prediction model offers genuine improvement over existing models used in clinical practice. Accurately powered studies are essential to confidently assess a new model's performance and to conduct robust comparisons against established reference models.

This work has a few limitations. First, while powerROC represents an important step toward making sample size calculation for model validation more accessible, its current functionality is limited to the AUROC, which quantifies discrimination. Future iterations of powerROC could incorporate additional performance metrics relevant to clinical prediction models, such as calibration measures, to provide researchers with a more comprehensive tool for study design. Second, we recognize that specifying a probability distribution in the absence of a pilot test set can be challenging for users, despite our efforts to simplify this process. Existing approaches often rely on specifying parameters for bivariate binormal distributions \cite{zhou2014statistical}, typically providing minimal guidance or intuitive tools to assist users. In contrast, powerROC offers a more user-friendly approach by incorporating default values based on empirical findings, providing clear explanations of the required parameters, and utilizing intuitive visualizations like contour plots to aid user understanding.

It is important to note that this work focuses specifically on external validation. Sample size considerations for model development represent a distinct, yet equally important, area of research \cite{riley2019minimum,riley2020calculating}. Future work could explore extending the principles and functionalities of powerROC to address the unique challenges of sample size determination during the model development phase. Continued research in these domains is crucial to further improve the accessibility and practicality of sample size calculation methods, ultimately promoting more rigorous and reliable prediction models in healthcare.

\subsection*{Conclusion}
This work underscores the importance of adequate statistical power in external validation studies of prediction models. powerROC, an open-source web tool, has been introduced to facilitate the sample size calculation for AUROC-based model comparisons. The tool offers a web-based interface and accommodates various study designs, enabling researchers to perform sample size calculations for external validations. Future development will focus on expanding powerROC’s functionality to encompass a wider range of performance metrics.

\subparagraph{Contributions}
FG: study ideation and design, created the powerROC web tool, manuscript first draft; RT: provided statistical expertise; JHC: study ideation, manuscript first draft. All authors reviewed and provided feedback on the manuscript.

\makeatletter
\renewcommand{\@biblabel}[1]{\hfill #1.}
\makeatother

\bibliographystyle{vancouver}
\bibliography{amia}  

\end{document}